\documentclass[reprint,twocolumn,showpacs,amsmath,amssymb,superscriptaddress,aps,prl]{revtex4-1}

\usepackage{graphicx}
\usepackage[english]{babel}
\usepackage{upgreek}	
\usepackage[update,prepend]{epstopdf}
\usepackage{etoolbox} 
\usepackage{color}
\usepackage{ulem}

\begin{document}

\title{Light-mediated collective atomic motion in an optical lattice coupled to a membrane}

\author{Aline Vochezer}\email{n\'{e}e Faber. Email: aline.faber@unibas.ch}
\affiliation{Department of Physics, University of Basel, Klingelbergstrasse 82, 4056 Basel, Switzerland}
\author{Tobias Kampschulte}
\altaffiliation[Present address: ]
{Institut f\"{u}r Quantenmaterie and Center for Integrated Quantum Science and Technology (IQ$^\textrm{ST}$), Universit\"{a}t Ulm, 89069 Ulm, Germany}
\affiliation{Department of Physics, University of Basel, Klingelbergstrasse 82, 4056 Basel, Switzerland}
\author{Klemens Hammerer}
\affiliation{Institute for Theoretical Physics and Institute for Gravitational Physics (Albert Einstein Institute), Leibnitz University Hannover, Callinstrasse 38, 30167 Hannover, Germany}
\author{Philipp Treutlein}\email{philipp.treutlein@unibas.ch}
\affiliation{Department of Physics, University of Basel, Klingelbergstrasse 82, 4056 Basel, Switzerland}

\begin{abstract}
We observe effects of collective atomic motion in a one-dimensional optical lattice coupled to an optomechanical system. 
In this hybrid atom-optomechanical system, the lattice light generates a coupling between the lattice atoms as well as between atoms and a micromechanical membrane oscillator. For large atom numbers we observe an instability in the coupled system, resulting in large-amplitude atom-membrane oscillations. 
We show that this behavior can be explained by light-mediated collective atomic motion in the lattice, which arises for large atom number, small atom-light detuning and asymmetric pumping of the lattice, in agreement with previous theoretical work. The model connects the optomechanical instability to a phase delay in the global atomic back-action onto the lattice light, which we observe in a direct measurement. 
\end{abstract}

\maketitle

Ultracold atoms in optical lattice potentials formed by interference of laser beams are a powerful system for many-body physics \cite{Bloc08}, quantum information science \cite{Jaksch05} and precision metrology \cite{Ludlow15}. 
In most optical lattice experiments the lattice light is far detuned from any atomic resonance, providing a conservative external potential with negligible back-action of the atoms onto the lattice light. 
The back-action is significantly enhanced when operating at moderate atom-light detuning in the tens of MHz to GHz regime \cite{Deutsch95,Birkl95,Weidemuller98} and at high atomic density. In this regime, the lattice light can mediate long-range interactions that couple the motion of atoms in different lattice potential wells. These interactions have been predicted to give rise to a variety of intriguing phenomena, ranging from spontaneous self-ordering and crystallization of light and atoms \cite{Ostermann14,Ostermann16} to the appearance of traveling wavelike collective oscillations of the atoms that can destabilize the entire lattice \cite{Asbo07,Asbo08}. 
Moreover, the back-action onto the lattice light can also be exploited to dynamically couple the atoms to other physical systems such as micromechanical oscillators \cite{Hung11,Treu14,Came11a,Joec14,Moll16}. Such hybrid atom-optomechanical systems offer new perspectives for ground-state cooling and quantum control of engineered mechanical structures \cite{Hamm10b,Voge13,Benn14,Hamm09a,Carmele14,Vogell15} and for studies of nonequilibrium quantum phase transitions \cite{Mann17}.

In the experiments reported here we observe effects of light-mediated collective atomic motion in an optical lattice in the context of building a hybrid system where the lattice light couples the atoms to a micromechanical membrane oscillator. The membrane acts like an additional ``super-atom'' of particularly high polarizability, enhancing collective effects in the entire system and providing a convenient way to directly detect the dynamics of the coupled system. For large numbers of atoms in the lattice we observe a dynamic instability of the coupled system, which can be explained in a model that takes light-mediated interactions of the atoms into account. These long-range atomic interactions also lead to an additional phase delay in the global atomic back-action onto the light field, which we observe in experiments. The phase delay can induce unstable behavior in the hybrid system if the atom-membrane coupling is large, even if the lattice itself is still stable. 
Our experiments show that light-mediated atom-atom interactions are significant even in free-space optical lattices, providing a way to study nonequilibrium many-body physics \cite{Asbo07,Asbo08,Ostermann14,Ostermann16,Mann17} that is complementary to experiments with atoms in optical cavities \cite{Rits13}.

\begin{figure}%
\includegraphics[width=\columnwidth]{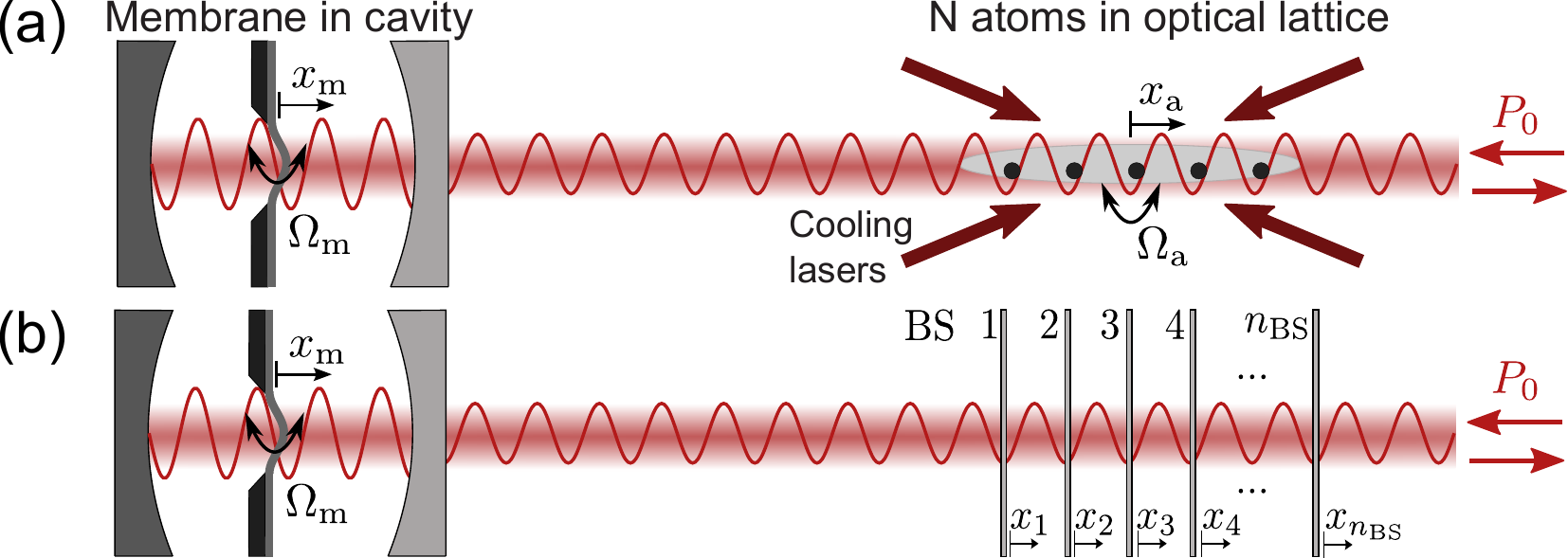}
\caption{(a) Optomechanical coupling scheme of atoms in an optical lattice and a micromechanical membrane oscillator in an optical cavity. (b) Modeling the atoms as beam splitters (BSs) allows to describe light-mediated collective motion of the atoms (see text). }
\label{fig:Fig1_setup}
\end{figure}

Our hybrid system is illustrated in Fig.~\ref{fig:Fig1_setup}(a). It consists of a Si$_3$N$_4$ membrane oscillator with mass $M=117\,$ng and vibration frequency $\Omega_\textrm{m}=2\pi \times 276\,$kHz in an optical cavity \cite{Jayi08a} and an ensemble of $N$ ultracold Rubidium atoms of mass $m$ in an optical lattice. The lattice is generated by a laser beam which also drives the membrane-cavity with a detuning $\Delta$ much smaller than the cavity linewidth $\kappa$ \cite{Supp17}. For large frequency detunings $\Delta_\textrm{LA}$ of the driving laser from the atomic transition and small atomic densities, light-mediated atom-atom interactions are negligible and the atoms oscillate with a frequency $\Omega_\textrm{a}\propto \sqrt{P_0}$ in the lattice potential wells, adjustable via the driving laser power $P_0$ \cite{Grim00}.
As predicted in~\cite{Hamm10b,Voge13,Benn14} and observed in~\cite{Came11a,Joec14}, radiation-pressure forces mediated by the lattice light couple the vibrations of atoms and membrane over a large distance. In absence of collective atomic effects all atoms couple equally to the membrane, resulting in a linear coupling of the membrane displacement $x_m$ to the atomic center of mass displacement $x_a$ with a coupling constant $g_N=|r_\textrm{m}|\Omega_\textrm{a}\sqrt{Nm\Omega_\textrm{a}/M\Omega_\textrm{m}}\,(2F/\pi)$, where $F=570$ is the cavity finesse and $r_\textrm{m}=0.41$ the membrane reflectivity ~\cite{Hamm10b, Voge13}.
The coupling mechanism exploits the fact that a displacement of the membrane induces a phase shift of the reflected light, which displaces the lattice potential wells. Conversely, a displacement of the atoms changes the optical power traveling towards the membrane and with this the radiation pressure force on the membrane. 
Additional cooling lasers applied to the atoms result in strong damping of the atomic motion at a rate $\Gamma_\textrm{a}\gg g_N$ and cool the atomic cloud to $4\,$mK.
In absence of atom-atom interactions the coupling then leads to sympathetic cooling of the membrane vibrations at a rate $\Gamma_\textrm{sym}=4\eta^2t^2g_\textrm{N}^2/\Gamma_\textrm{a}$ (for resonant coupling $\Omega_\textrm{a}=\Omega_\textrm{m}$). Here $\eta \approx 1$ is the incoupling efficiency into the optical cavity and $t=0.71$ the amplitude transmission of the optical path between atoms and membrane~\cite{Joec14, Came11a, Voge13}. This has been used in the experiments of ref.~\cite{Joec14} to cool a membrane oscillator from room temperature to 0.7\,K using the atoms as coolant.

\begin{figure}%
\includegraphics[width=\columnwidth]{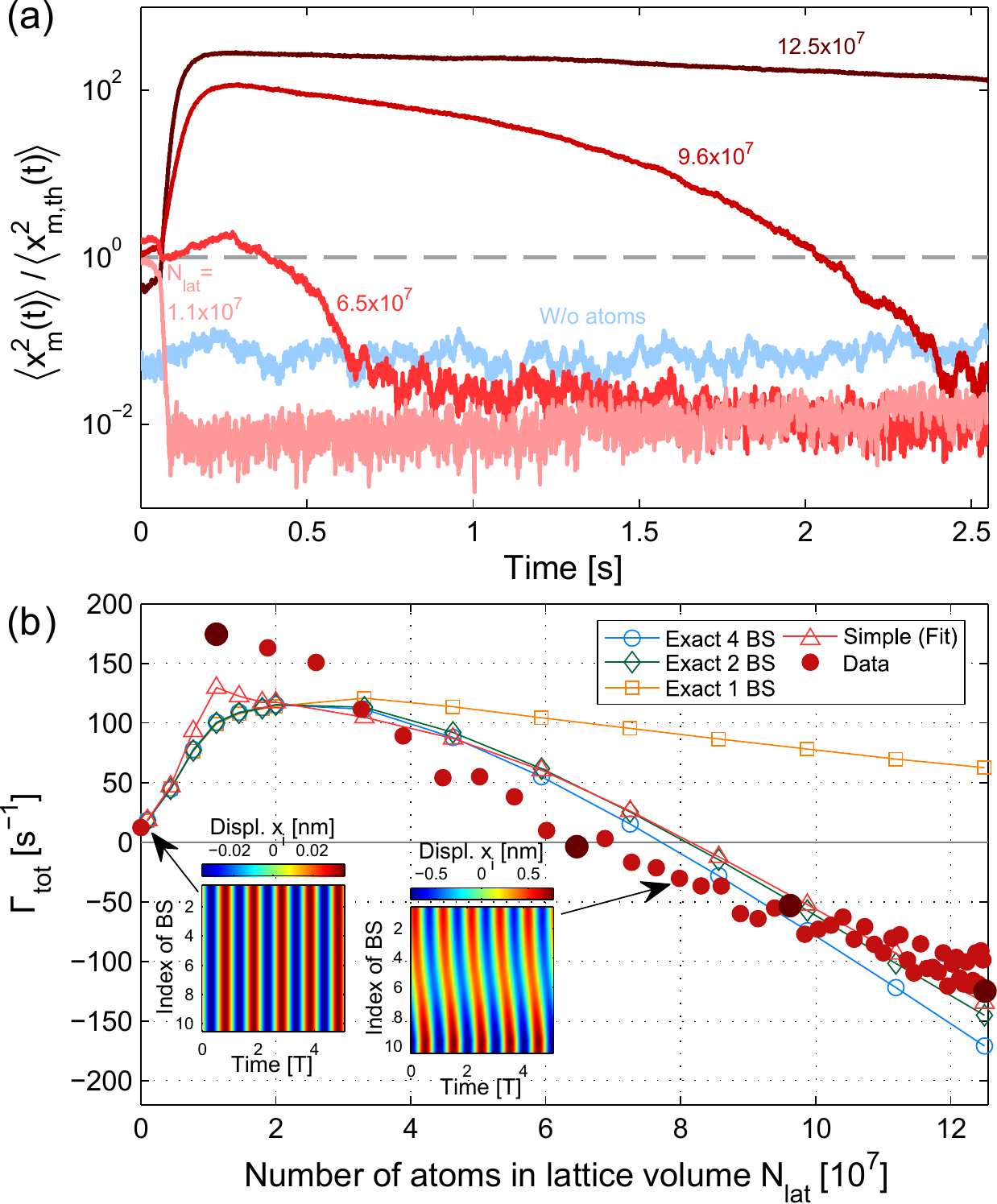}
\caption{{Observation of self-oscillations.} (a) Evolution of $\langle x_\textrm{m}^2(t) \rangle$ with time for different $N_\textrm{lat}$. Red traces: Lattice is ramped up at $t=50\,$ms. Blue trace: Lattice is running at $P_0=3.4\,$mW continuously with $N_\textrm{lat}=0$. Dashed gray: Room temperature level. (b) Total membrane damping rate $\Gamma_\textrm{tot}$ versus $N_\textrm{lat}$. Filled red circles: Data extracted from traces as in (a). Larger red circles: Data points of traces in (a). Empty blue circles: Numerical simulation with exact model with 4 beam splitters (BS). Empty green diamonds: As blue, but with 2 BS. Empty orange squares: As blue and green, but with only 1 BS. Empty red triangles: Simulation of linearized model with 2 BS fitted to the data. For all four curves, $\Gamma_\textrm{a}=233\,\textrm{s}^{-1}$ and $\alpha=0.11$. Insets: Simulated displacement $x_\textrm{i}$ of an array of 10 BS as a function of time for $0.3 \times 10^7$ (left) and $8\times 10^7$ (right) atoms. }
\label{fig:Fig2_Selfosci}
\end{figure}

The sympathetic cooling measurements of ref.\,\cite{Joec14} were performed with large light-atom detuning $\Delta_\textrm{LA}=-2\pi\times 8\,$GHz from the $F=2 \leftrightarrow F^\prime =3$ transition of the $^{87}$Rb D$_2$ line at $\lambda=780\,$nm. Here we focus on small $\Delta_\textrm{LA}\approx -2\pi\times 1\,$GHz where the coupled dynamics becomes drastically different. An instability occurs at large atom numbers, where the membrane amplitude starts to grow exponentially because the total membrane damping rate $\Gamma_\textrm{tot}=\Gamma_\textrm{m}+\Gamma_\textrm{opt}+\Gamma_\textrm{sym}$ changes sign from positive to negative. In $\Gamma_\textrm{tot}$ we include the intrinsic membrane damping rate $\Gamma_\textrm{m}=0.96\,\textrm{s}^{-1}$ and standard cavity-optomechanical damping at rate $\Gamma_\textrm{opt}=10.6\,\textrm{s}^{-1}$~\cite{Aspe13}, which arises from the small red laser-cavity detuning $\Delta=-0.06\,\kappa$.
We observe this instability in experiments where we detect the membrane amplitude with an additional detection beam and vary the number of atoms in the lattice volume $N_\textrm{lat}$~\cite{Supp17}. The number of resonantly coupled atoms $N$ is smaller than $N_\textrm{lat}$ because of the inhomogeneous transverse lattice profile. As in \cite{Joec14}, we estimate $N= \frac{\pi \Gamma_\textrm{a}}{2\Omega_\textrm{m}} N_\textrm{lat}$. During the preparation of the atomic ensemble the lattice is operating at low driving laser power $P_0=0.1\,$mW and the atoms do not couple to the membrane because $\Omega_\textrm{a}\ll \Omega_\textrm{m}$. At the start of the coupling experiment the lattice is ramped up in 10\,ms to $P_0=3.4\,$mW, which tunes the atomic oscillation frequency into resonance with the membrane frequency.
The red traces in Fig.~\ref{fig:Fig2_Selfosci}(a) exemplarily show how the mean square membrane displacement $\langle x_\textrm{m}^2(t) \rangle$ subsequently evolves with time for different $N_\textrm{lat}$. 
For the smallest $N_\textrm{lat}$ we observe strong damping of the membrane motion resulting in a steady state value below the optomechanical cooling level (blue curve). This is the sympathetic cooling effect observed in~\cite{Joec14}. Subsequent atom loss from the ensemble on the timescale of several seconds reduces the cooling effect as expected from $\Gamma_\textrm{sym}\sim N$. For the next larger $N_\textrm{lat}$ the membrane amplitude decreases only after some atoms have been lost. For even larger $N_\textrm{lat}$ the system becomes unstable. Now the membrane amplitude increases after the turn-on and performs limit cycle oscillations at a large amplitude $\langle x_\textrm{m}^2(t)\rangle/\langle x_\textrm{m,th}^2(t)\rangle\approx 100$ before it slowly decreases when atoms are lost. Here $\langle x_\textrm{m,th}^2(t)\rangle^{1/2}$ is the room temperature thermal amplitude of the membrane. Thus, we find that at large atom number the presence of the atoms induces an instability, corresponding to negative $\Gamma_\textrm{tot}$. This behavior is not predicted by previous theoretical descriptions of the atom-membrane coupling \cite{Hamm10b,Voge13}, which neglect collective atomic effects. 
From the slopes of $\langle x_\textrm{m}^2(t) \rangle$ after the turn-on of the lattice, we can extract $\Gamma_\textrm{tot}$, which is plotted in Fig.~\ref{fig:Fig2_Selfosci}(b) against $N_\textrm{lat}$ (filled red circles). 
 
We can model the observed behavior if we take light-mediated interactions between the atoms in different lattice wells into account. Following~\cite{Asbo08}, we model the combined effect of the $N_\textrm{BS}$ atoms in each potential well as a thin beam splitter (BS) as illustrated in Fig.~\ref{fig:Fig1_setup}(b) with reflection and transmission coefficients given by the dimensionless atomic polarizability density $\zeta= \frac{\Gamma}{-\Delta_\textrm{LA}}\frac{N_\textrm{BS} \lambda^2}{4\pi \sigma_\textrm{L}}$, where $\Gamma$ is the natural linewidth of the atomic transition and $\sigma_\textrm{L}$ the transverse area of the laser beam. The imaginary part of $\zeta$ is omitted since $|\Delta_\textrm{LA}|\gg \Gamma$. We calculate the forces on the $n_\textrm{BS}=N/N_\textrm{BS}$ atomic BSs and the membrane using the transfer matrix method. The detailed model is presented in~\cite{Supp17}, where we apply the theory of~\cite{Asbo08} to our system. With this model we perform numerical simulations of the coupled dynamics and extract a theoretical value for $\Gamma_\textrm{tot}$.
The model predicts the instability at large $N$ and that anharmonicities in the atomic potential stop the exponential growth of the membrane amplitude and stabilize the limit cycle oscillation. 

The blue (green) circles (diamonds) in Fig.~\ref{fig:Fig2_Selfosci}(b) show the result of numerical simulations with the atoms distributed over four (two) atomic BSs. The red triangles are the result of an analytical analysis (see below) and the orange squares show the damping rate for one atomic BS, i.e. without collective effects. The numerical simulations have been performed for a reasonable $\Gamma_\textrm{a}=233\,\textrm{s}^{-1}$, and $N_\textrm{lat}$ has been scaled by a factor $\alpha=0.11$ for the simulation to match the data. This is plausible as the temperature of the atomic cloud ($ \approx 4\,$mK) is larger than the depth of the coupling lattice ($ \approx 500\,\mu$K) so that not all atoms are trapped. Both traces (blue and green) do not exactly reproduce the data. This can be due to the fact that the model of the atomic ensemble is greatly simplified. As we operate the system when the lattice is overlapped with a large magneto-optical trap~\cite{Supp17}, we do not have direct access to the atoms taking part in the coupling. However, the main features such as the initial linear increase of $\Gamma_\textrm{tot}$ and the subsequent decrease leading to negative damping are confirmed by the model and the numbers match roughly. If we replace the membrane by a fixed mirror, we also simulate unstable behavior as in~\cite{Asbo07,Asbo08} but for larger $N_\textrm{lat}$.

The traces with four and two BSs differ only slightly, whereas the simulation with only one atomic BS (orange squares) does not show the instability~\footnote{The curve with one BS saturates when $g_N\approx \Gamma_\textrm{a}$ and decreases for larger $N_\textrm{lat}$ because also a single BS modifies the light field configuration. However $\Gamma_\textrm{tot}$ remains positive for all $N_\textrm{lat}$ in this case.} indicating that coupled motion of atoms in different lattice wells plays an essential role. We observe that the behavior quickly converges for more than two BSs suggesting that only a few collective atomic modes are relevant. The insets in Fig.~\ref{fig:Fig2_Selfosci}(b) show exemplarily how the displacements $x_\textrm{i}$ of ten BSs evolve as a function of time for $0.3 \times 10^7$ atoms and $8 \times 10^7$ atoms. For the small atom number all BSs move in phase and do not interact so that only their center of mass motion couples to the membrane. For the large atom number a traveling wavelike collective oscillation appears. In this case more than one collective atomic mode must take part in the coupling.
Given the fast convergence for more than two BS we have a closer look at the simplest model, the membrane coupled to a stack of two BSs.
For this two-BS model we linearize the radiation pressure forces around the steady state positions of the BSs and the membrane and expand the linear coefficients up to third order in $\zeta$. This model enables us to describe the onset of instability, but does not cover the regime of limit cycles.
We find the following linear equations of motion for the displacement of the membrane $x_\textrm{m}$ and the two atomic BSs $x_\textrm{1}$ and $x_\textrm{2}$:
{\allowdisplaybreaks
\begin{eqnarray}
\label{eqn:2BSmodel}
\ddot{x}_\textrm{m}&=&-\Gamma_\textrm{m}'\dot{x}_\textrm{m}-\Omega_\textrm{m}^2 x_\textrm{m}+k_\textrm{mm}x_\textrm{m}+k_\textrm{m1}x_\textrm{1}+k_\textrm{m2}x_\textrm{2}\,, \nonumber \\
\ddot{x}_\textrm{1}&=&-\Gamma_\textrm{a}\dot{x}_\textrm{1}+k_\textrm{1m}x_\textrm{m}+k_\textrm{11}x_\textrm{1}+k_\textrm{12}x_\textrm{2}\,, \nonumber \\
\ddot{x}_\textrm{2}&=&-\Gamma_\textrm{a}\dot{x}_\textrm{2}+k_\textrm{2m}x_\textrm{m}+k_\textrm{21}x_\textrm{1}+k_\textrm{22}x_\textrm{2}\,,
\end{eqnarray}}
with $\Gamma_\textrm{m}'=\Gamma_\textrm{m}+\Gamma_\textrm{opt}$ and coefficients

{\allowdisplaybreaks
\begin{align}
k_\textrm{mm} &= \frac{N m}{2M}\Omega_\textrm{a}^2 R(-2+10\nu)f^2\,,&&\nonumber \\
k_\textrm{m1} &= \frac{N m}{2M}\Omega_\textrm{a}^2 R(1-9\nu)f\,,&&\nonumber \\
k_\textrm{m2} &= \frac{N m}{2M}\Omega_\textrm{a}^2 R(1-\nu)f\,,&&\nonumber \\
k_\textrm{1m} &= \Omega_\textrm{a}^2(1-\nu)f\,,& k_\textrm{2m} &= \Omega_\textrm{a}^2(1-9\nu)f\,, \nonumber \\
k_\textrm{11} &= \Omega_\textrm{a}^2(-1+\nu)\,,& k_\textrm{21} &= \Omega_\textrm{a}^28 \nu\,,\nonumber \\
k_\textrm{12} &= 0\,, & k_\textrm{22} &= \Omega_\textrm{a}^2(-1+\nu)\,.
\end{align}}

Here $R=\eta t^2$ is the lattice amplitude reflection coefficient, $f=2 |r_\textrm{m}|\frac{2 F}{\pi}$ the cavity enhancement factor and $\nu=\frac{\mathcal{A}^2\zeta^2}{8}$ a dimensionless parameter which depends on the polarizability density $\zeta \propto N$ of a single atomic BS and the lattice asymmetry $\mathcal{A}=(1-R^2)/R$. The parameter $\nu$ describes the effect of collective atomic motion in leading order of $\zeta$, and $\nu=0$ recovers the case of non-interacting atoms. In the experiments $\mathcal{A}$ is fixed, so that $\nu$ scales with $N$.
A numerical simulation of these simplified equations of motion (red triangles in Fig.~\ref{fig:Fig2_Selfosci}(b)) reproduces the exact result with two BS (green diamonds) as expected. For $\nu\ll 1$, i.e. small $N$, both BSs couple to the membrane equally and move independently of each other. We can then rewrite the equations of motion as a coupling between the membrane displacement $x_\textrm{m}$ and the atomic center of mass displacement $x_\textrm{a}=(x_\textrm{1}+x_\textrm{2})/2$ reproducing the result of ref. \cite{Joec14, Voge13}. However, for larger $N$ (large $\nu$) the atoms interact with each other and the membrane does not couple to a single atomic mode any more. Note that the dynamics described by the set of Eqs.~\ref{eqn:2BSmodel} is non-conservative ($k_{21}\neq k_{12}=0$), a consequence of the cascaded nature of the system.

\begin{figure}%
\includegraphics[width=\columnwidth]{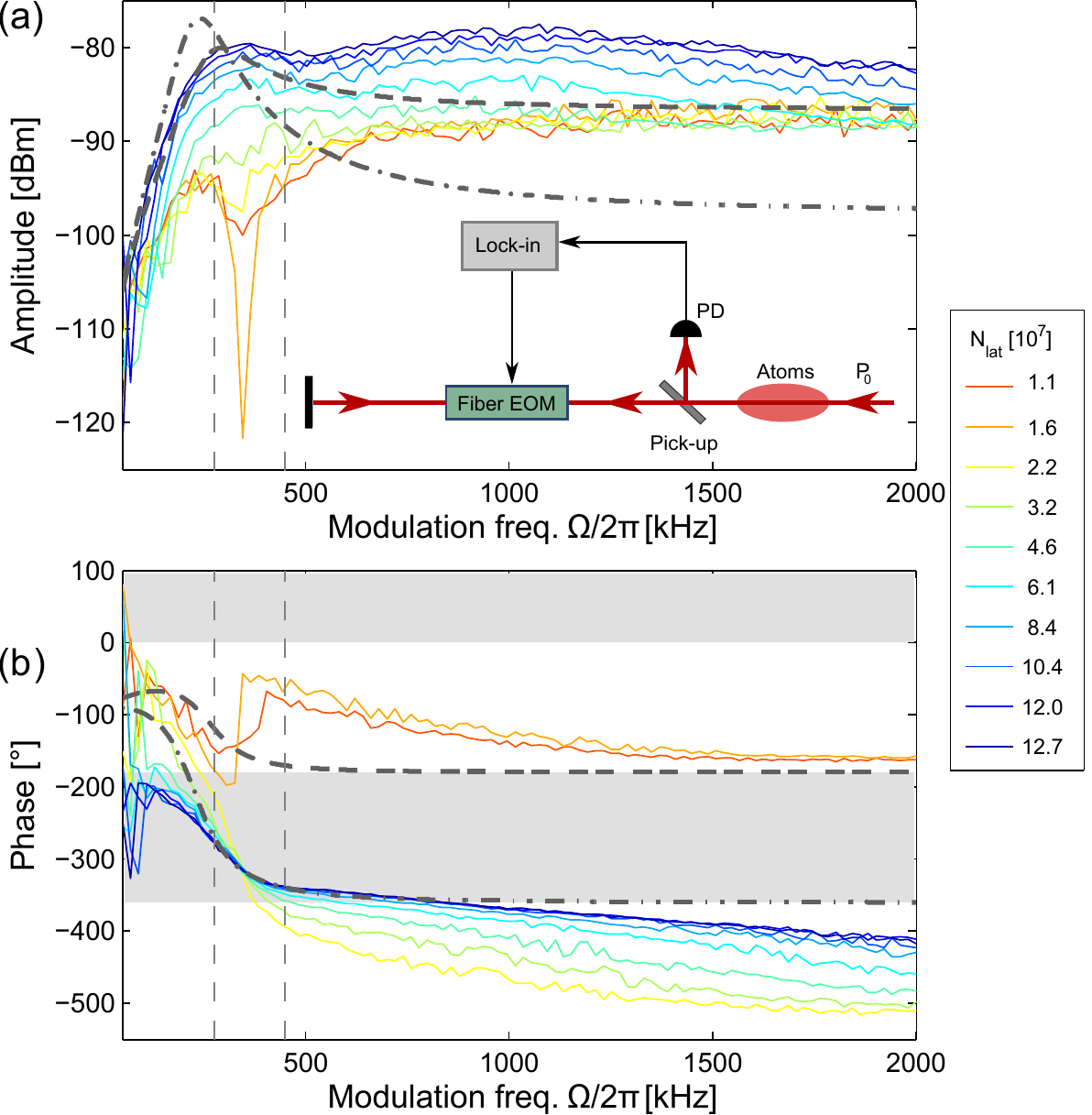}
\caption{{Atomic back-action on the lattice light.} (a) Amplitude (electrical power) and (b) phase of atomic back-action versus modulation frequency $\Omega$. Dashed vertical lines: Membrane frequency $\Omega_\textrm{m}/2\pi= 276\,$kHz and frequency of the atoms in the center of the trap $\Omega_\textrm{a}(0)/2\pi=450\,$kHz. Shaded areas: region in which the coupled system can become unstable if the coupling is strong enough. Inset in (a): Measurement setup. PD:photodiode. Thick dashed (dashed-dotted) lines: Behavior expected from one-BS-model (two-BS-model) for $N=3\times 10^8$, $\Omega_\textrm{a}=2\pi\times 275\,$kHz, $\Gamma_\textrm{a}=2\pi\times 150\,$kHz and $R=0.06$. The modeled amplitudes are scaled down by 43\,dB to adjust to the data. }
\label{fig:Fig3_BackactionMeas}
\end{figure}

The instability can also be understood in a feedback picture:
The coupled atom-membrane motion becomes instable if a signal traveling from the membrane to the atoms and back experiences a phase delay of $360^\circ$ and a  loop-gain larger than unity~\cite{Fran14}. A system of only two coupled harmonic oscillators, such as the membrane coupled to one atomic oscillator (e.g. the atomic center of mass motion), cannot become unstable as the maximum phase delay collected in one round trip stays below $2 \times 180^\circ=360^\circ$. If an additional harmonic oscillator e.g. in form of a second collective atomic mode takes part in the coupling, the atomic back-action onto the laser beam traveling towards the membrane can be delayed by more than $180^\circ$, providing a necessary condition for instability of the coupled system. 

To directly observe this phase delay, we performed experiments in which the phase shift induced by the membrane was mimicked by a fiber electro-optic modulator (EOM) and the atomic back-action onto the lattice power was detected with a photodiode as depicted in the inset of Fig.~\ref{fig:Fig3_BackactionMeas}(a)~\cite{Supp17}. Fig.~\ref{fig:Fig3_BackactionMeas}(a) and (b) show how amplitude and phase of the photodiode signal evolve as a function of the EOM modulation frequency $\Omega$ for different $N_\textrm{lat}$. The dashed (dashed-dotted) traces show the expected behavior for the one-BS-model used in ref.~\cite{Joec14, Voge13} (the two-BS-model), see \cite{Supp17}. The one-BS-model predicts a maximum phase delay of $180^\circ$. In contrast, for large atom numbers the data and the two-BS-model show phase delays $> 180^\circ$ indicating that the one-BS model is not sufficient to describe the system and showing that the coupled atom-membrane motion can indeed become unstable. For the theory curves the inhomogeneously broadened atomic ensemble has been modeled with all atoms ($N=N_\textrm{lat}$) and an increased, inhomogeneously broadened atomic linewidth $\Gamma_\textrm{a}$ \cite{Supp17} in contrast to Fig.~\ref{fig:Fig2_Selfosci}(b) where only the resonant atoms are taken into account. Insufficient knowledge of the exact properties of the atoms in the lattice makes a more precise modeling of the atomic back-action difficult. The great simplification in modeling the atomic ensemble as well as uncertainties in the signal calibration lead to a discrepancy in the signal amplitude heights between data and theory. Still, one- and two-BS-model show a drastic difference in the phase behavior for realistic parameters.

An additional phase delay enters into our system from the propagation time delay between atoms and membrane $\tau_\textrm{prop}=30\,$ns and the finite response time of the cavity $\tau_\textrm{cav}=0.6\,$ns. However, these delays are of minor importance for the stability of our system, which was confirmed by varying the path length between atoms and membrane~\cite{Supp17}.

In summary, we observed for the first time light-mediated atom-atom interactions in a free space optical lattice giving rise to collective atomic oscillations and lattice instabilities. In our experiment these effects are enhanced by coupling the atoms to a distant dielectric membrane oscillator, which at the same time serves as a sensitive probe for the light-mediated collective atomic motion. The instabilities and collective dynamics in this hybrid 
system are described well by a model adapted from Asboth et al.~\cite{Asbo08}.

Our experiment shows that substantial light-mediated atom-atom interactions can arise in free-space optical lattices in a regime of large atom numbers, moderate atom-light detuning and asymmetric driving of the lattice. This offers new possibilities for the study of many-body physics, such as the spontaneous crystallization of atoms and light into a structure that features phonon-like excitations and bears similarities to a supersolid \cite{Ostermann16}.
Moreover, our results are relevant for the development of hybrid atom-optomechancial systems in the quantum regime \cite{Hung11,Treu14,Came11a,Joec14,Moll16}. The configuration studied in our experiment has been proposed for ground-state cooling of mechanical oscillators in regimes where purely optomechanical techniques fail \cite{Hamm10b,Voge13,Benn14}. Variants of the setup have been suggested for the generation of non-classical vibrational states of mechanical oscillators \cite{Carmele14,Vogell15}. In both cases, light-mediated atom-atom interactions have to be taken into account. Finally, such interactions could be harnessed to study nonequilibrium quantum phase transitions \cite{Mann17}.


\begin{acknowledgments}
We thank Niels L\"{o}rch, Christoph Bruder, Andreas J\"{o}ckel and Dan Stamper-Kurn for useful discussions. This work was supported by the ERC project MODULAR.
\end{acknowledgments}

\section{Appendix}

\subsection{Experimental setup}

\begin{figure}%
\includegraphics[width=\columnwidth]{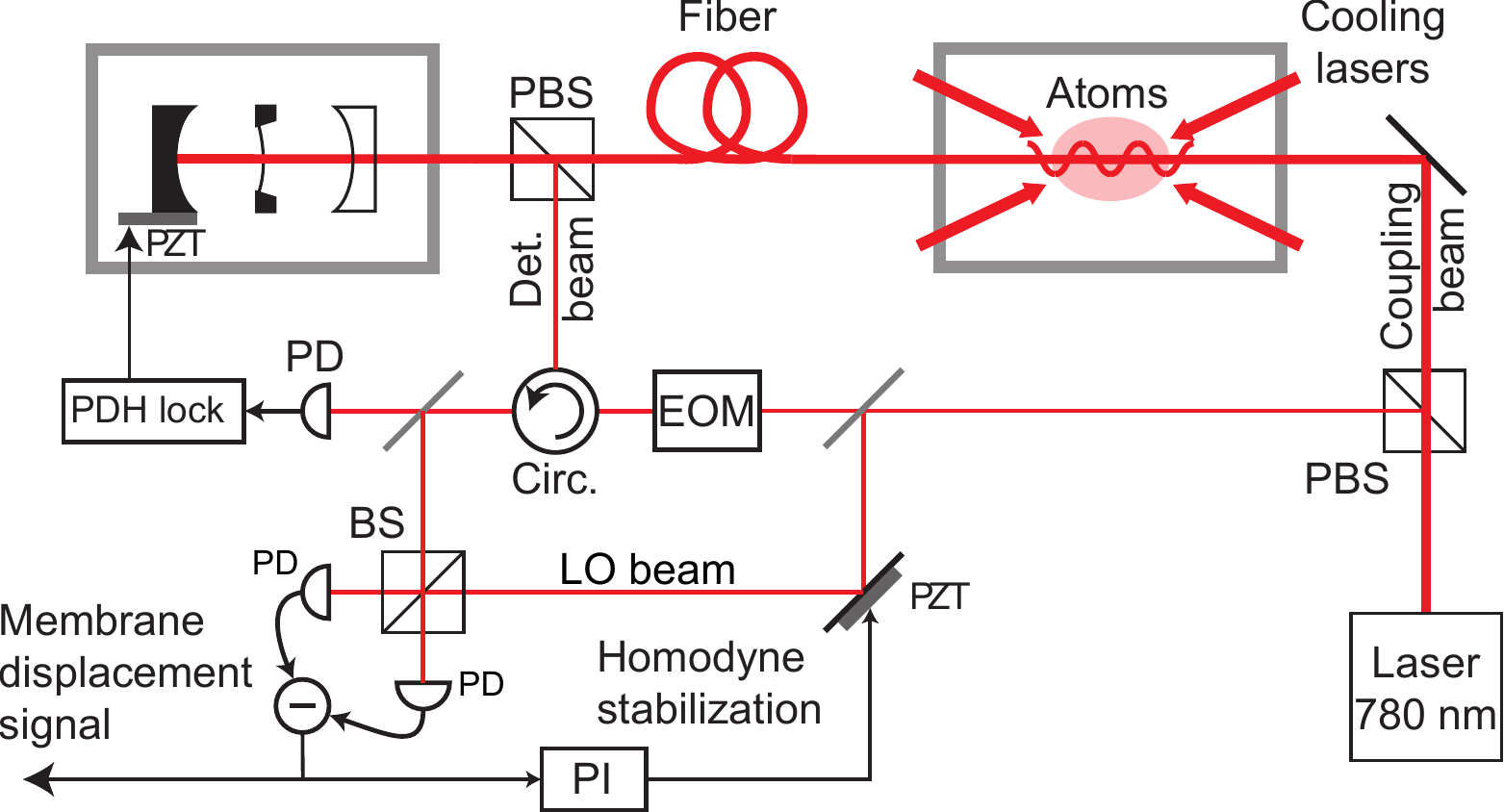}
\caption{Experimental setup. The membrane-cavity system and the atoms are residing in separate vacuum chambers (grey boxes) connected by a coupling laser beam via an optical fiber. A $100\,\mu$W readout and detection beam is split off the coupling beam at a polarizing beam splitter (PBS) and coupled to the cavity with orthogonal polarization. Using the Pound-Drever-Hall (PDH) method an error signal is created, which is used to stabilize the laser-cavity detuning. The major part of the reflected detection light is overlapped with a local oscillator beam (LO) on a beam splitter (BS) and detected by photo diodes (PD), realizing a homodyne detection scheme. From the PD signal the power spectral density of the membrane displacement $S_x(\Omega)$ is extracted.  EOM: electro-optic modulator, PZT: piezo electric transducer. }
\label{fig:Supple0_Setup}
\end{figure}

The experiments were performed with the system sketched in Fig.~\ref{fig:Supple0_Setup}. The mechanical oscillator is a 41\,nm thin and $1.5\,$mm$\times 1.5\,$mm wide Si$_3$N$_4$ membrane. It oscillates like a square drum with fundamental mode frequency $\Omega_\textrm{m}=2\pi \times 276\,$kHz, damping rate $\Gamma_\textrm{m}=0.96\,\textrm{s}^{-1}$ and effective mass $M=117\,$ng. At our wavelength of $\lambda=780\,$nm the membrane is semi-transparent with $r_\textrm{m}=0.41$. It resides inside a single-sided optical cavity of linewidth $\kappa=2\pi\times290\,$MHz and finesse $F=570$ creating a membrane-in-the-middle system with a single-photon optomechanical coupling strength $g_0=690\,\textrm{s}^{-1}$~\cite{Aspe13}. The membrane is placed at a position near but not exactly on the slope of the intracavity standing wave where the optomechanical and atom-membrane coupling strengths are reduced by a factor $0.63$ from their maximum values. The coupling beam with power $P_0$ enters the system from the right. It travels through the atomic ensemble and is reflected off the membrane-cavity system ($\eta \approx 1$). We operate the cavity at a small red detuning $\Delta = -0.06 \kappa$ to avoid the optomechanical parametric instability \cite{Aspe13} leading to weak optomechanical damping $\Gamma_\textrm{opt}=10.6\,\textrm{s}^{-1}$. The detection system illustrated in Fig.~\ref{fig:Supple0_Setup} allows to monitor the membrane displacement spectrum $S_x(\Omega)$ with a spectrum analyzer. The coupling beam is red-detuned by $\Delta_\textrm{LA}=-2\pi\times 960\,$MHz from the $F=2 \leftrightarrow F^\prime =3$ transition of the $^{87}$Rb D$_2$ line at $780\,$nm. It creates an optical lattice potential for the atoms, in which they oscillate with frequency $\Omega_\textrm{a}\propto \sqrt{P_0/\Delta_\textrm{LA}}$ if atom-atom interactions are irrelevant. The lattice beam waist at the position of the atoms is $w_0=280\,\mu$m and the amplitude transmission between atoms and membrane-cavity system is $t=0.71$ so that there is a power imbalance between the counter propagating lattice beams. Ultracold atoms are loaded into the lattice by overlapping it with a magneto-optical trap (MOT) as described in the following section. 

\subsection{Preparation and probing of the atomic ensemble}

To prepare an atomic ensemble with a defined number of atoms in the lattice volume $N_\textrm{lat}$ we overlap the lattice with a magneto-optical trap (MOT) and load the MOT for a variable time from a two-dimensional MOT (2D MOT). At the start of the coupling experiment the 2D MOT is switched off. The atom number then decreases slowly on the time scale of several seconds much slower than the atom-membrane dynamics. As the temperature of the MOT ($\approx 4\,$mK) is larger than the depth of the coupling lattice ($\approx 500\,\mu$K) not all atoms are trapped in the lattice potential. We use this unknown fraction as a fit parameter for the fit of the linearized two-BS model to the data (empty red triangles in Fig.~2(b) of the main text). The number of resonantly coupled atoms $N$ is further reduced by a factor $\frac{\pi \Gamma_\textrm{a}}{2\Omega_\textrm{m}}$ due to the inhomogeneous transverse lattice profile \cite{Joec14}.

We determine $N_\textrm{lat}$ by absorption imaging with a probe beam which is mode matched with the lattice beam.

\subsection{Calibration of the membrane displacement}
We calibrate the membrane displacement with the coupling beam turned off.
In our setup optomechanical damping effects from the weak detection beam are finite but small. Therefore and because large signals are investigated, we neglect this small cooling effect and calibrate the displacement axis in Fig.~2(a) by setting the signal in presence of the weak detection beam equal to the expected membrane signal at room temperature.

\subsection{Determination of $\Gamma_\textrm{opt}$}
The optomechanical damping rate $\Gamma_\textrm{opt}$ is determined from the temporal average of the membrane displacement in absence of atoms (blue trace in Fig.~2(a)) using \cite{Aspe13}
\begin{equation}
\langle x_\textrm{m}^2(t) \rangle=\langle x_\textrm{m,th}^2(t) \rangle\frac{\Gamma_\textrm{m}}{\Gamma_\textrm{m}+\Gamma_\textrm{opt}}\,,
\end{equation}
which describes optomechanical cavity cooling in the classical limit.

\subsection{Atom-membrane coupling with atoms in different potential wells}
\label{seq:MultiBSModel}

\begin{figure}%
\includegraphics[width=\columnwidth]{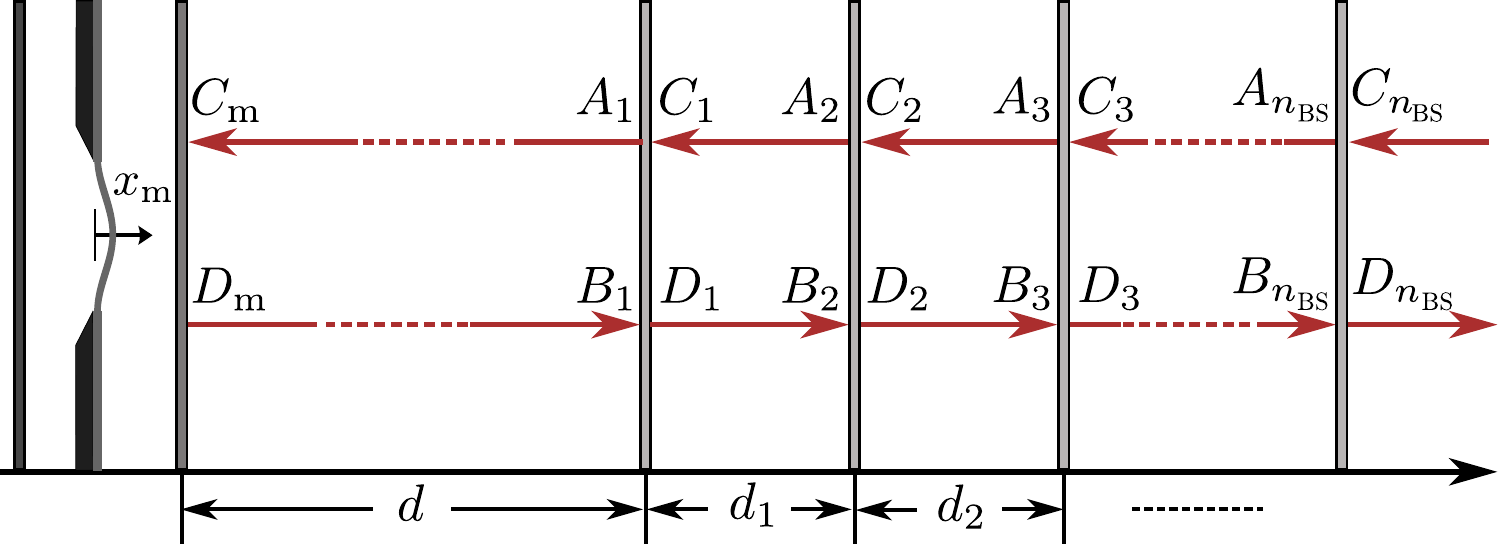}
\caption{{Model of atom-membrane system. The membrane resides inside an Fabry-Perot cavity. The atoms in different lattice wells are modeled as thin beamsplitters. The arrows symbolize the light field between the elements. A single laser beam enters the system from the right. As the back mirror of the membrane-cavity-system has a high reflectivity, no light leaves the system towards the left.}}
\label{fig:Supple1_ExactModel}
\end{figure}

For an extended modeling of the atom-membrane coupling including interactions between atoms in different potential wells we refer to the system sketched in Fig.~\ref{fig:Supple1_ExactModel}. 
The membrane is placed inside an asymmetric Fabry-Perot cavity. The back mirror has close to unity reflectivity but there can be losses in the input coupling so that a fraction $\eta\leq 1$ of the light is reflected back from the cavity system. The lattice laser drives the cavity close to resonance, maintaining a small red detuning $|\Delta|\ll \kappa$ to avoid the parametric instability on the blue side of the cavity resonance. If the membrane moves around its steady state position, it imprints a phase shift of $\Phi=4G x_\textrm{m}/\kappa$ onto the outgoing light \citep{Aspe13}, where $G=-\frac{\mathrm{d}\omega_\textrm{c}}{\mathrm{d}x_\textrm{m}}$ is the optomechanical coupling strength and $\omega_\textrm{c}$ the empty cavity resonance frequency \citep{Aspe13}. The field amplitudes $C_\textrm{m}$ and $D_\textrm{m}$ are therefore connected by
\begin{equation}
D_\textrm{m}=\eta e^{i\Phi}C_\textrm{m}\,.
\end{equation}
The radiation pressure force $F_\textrm{m}$ on the membrane in a membrane-in-the-middle (MIM) system is given by the power going into the cavity \citep{Aspe13}
\begin{equation}
\label{eqn:ForceMembrane}
F_\textrm{m}= \frac{4G}{\omega_\textrm{c} \kappa} P_\textrm{in}\,,
\end{equation}
where $P_\textrm{in}=\sigma_\textrm{L} \frac{ \epsilon_0 c |\eta C_\textrm{m}|^2}{2} $ with laser beam cross section $\sigma_\textrm{L}$, vacuum permittivity $\epsilon_0$ and speed of light $c$.

The atomic ensemble is modeled as an array of $n_\textrm{BS}$ thin sheets of polarizable material with polarizability \cite{Asbo08}
 \begin{equation}
 \label{eqn:PolarizabiltyAtom}
 \alpha=  \frac{\Gamma}{(-\Delta_\textrm{LA})}\frac{1}{4\pi^2}\epsilon_0\lambda^3\,,
 \end{equation} 
where $\Gamma$ is the natural linewidth and $\lambda$ the wavelength of the optical transition. Note that the expression is a factor $2/3$ smaller than the two-level result in \citep{Asbo08} due to the line strength of the $^{87}$Rb D$_2$ line. Further note that we omitted the imaginary part of the polarizability as it is negligible for $|\Delta_\textrm{LA}|\gg \Gamma$.
These thin sheets act as beam splitters (BSs) with transfer matrix $M$ which connects the fields left of each BS to the fields right of the BS \citep{Asbo08}
\begin{equation}
\label{eqn:MatrixFormulation}
\begin{bmatrix}A_i\\B_i\end{bmatrix}= M \begin{bmatrix}C_i\\D_i\end{bmatrix}=
\begin{bmatrix} 
									1+i\zeta & i\zeta \\
									-i \zeta & 1-i \zeta			
\end{bmatrix}
\begin{bmatrix}C_i\\D_i\end{bmatrix}\,,
\end{equation}
where $\textrm{i}= 1....n_\textrm{BS}$. Here $\zeta=k\eta \alpha/2\epsilon_0$ is the dimensionless polarizability density and $\eta=N_\textrm{BS}/\sigma_\textrm{L}$ the area density of the atoms in the sheet, with $\sigma_\textrm{L}$ being the transverse mode area of the beam and $N_\textrm{BS}=N/n_\textrm{BS}$ the number of atoms in one BS and in the mode volume.
The beamsplitters are separated by distances $d_\textrm{i}$ corresponding to the free-space transfer matrices 
\begin{equation}
\label{eqn:FreeSpaceProp}
M_{d,\textrm{i}}=
\begin{bmatrix} 
									e^{ikd_\textrm{i}}  & 0 \\
									0 			&e^{-ikd_\textrm{i}}  			
\end{bmatrix}\,,
\quad \textrm{i}= 1....n_\textrm{BS}\,.
\end{equation} 
Given an incoming plane wave from the right with amplitude $C(x)=C_0 \exp(-ikx)$ the field amplitudes at each position of the system can be calculated via the transfer matrices.
From the field amplitudes on the left and the right side of the ith BS the radiation pressure force on the BS can be calculated 
\begin{equation}
\label{eqn:ForceBS}
F_\textrm{i}=\frac{\epsilon_0 \sigma_\textrm{L}}{2}(|A_i|^2+|B_i|^2-|C_i|^2-|D_i|^2)\,.
\end{equation}
With the radiation pressure forces on the membrane and the ith BS we can write down the equations of motion for the membrane displacement $x_\textrm{m}$ and the displacement of the ith BS $x_\textrm{i}$ ($i=1...n_\textrm{BS}$)
\begin{eqnarray}
\label{eqn:eom}
M\ddot{x}_\textrm{m}&=&-M(\Gamma_\textrm{m}+\Gamma_\textrm{opt})\dot{x}_\textrm{m}-M\Omega_\textrm{m}^2 x_\textrm{m}+F_\textrm{m}\,, \nonumber \\
N_\textrm{BS}m\ddot{x}_\textrm{i}&=&-N_\textrm{BS}m\Gamma_\textrm{a}\dot{x}_\textrm{i}+F_\textrm{i}\,.
\end{eqnarray}
Here we added damping of the membrane with the intrinsic membrane damping rate $\Gamma_\textrm{m}$ and the cavity-optomechanical damping rate $\Gamma_\textrm{opt}$ and damping of the atomic motion due to laser cooling at rate $\Gamma_\textrm{a}$. The second term in the membrane equation describes the restoring force from the clamping of the membrane to its frame.

With these equations of motion the dynamics of the coupled atom-membrane system with $N$ atoms distributed over $n_\textrm{BS}$ BSs can be simulated numerically.
At the start of the simulation we displace the membrane slightly from its steady state position $x_\textrm{m}^\textrm{st}=F_\textrm{m,0}/M \Omega_\textrm{m}^2$. The steady state force on the membrane $F_\textrm{m,0}$ is given via Eq.~\ref{eqn:ForceMembrane} by the radiation pressure force of the ingoing power $P_\textrm{in}^\textrm{st}=\sigma_\textrm{L}\epsilon_0 c \eta^2 t^2 |C_0|^2/2$. Here $t$ is the amplitude transmission between atoms and membrane-cavity system.
Initially, the atoms are placed at their steady state positions.
To determine these steady state positions $x_\textrm{i}^\textrm{st}$, the phase reflected of the MIM system with the membrane at $x_\textrm{m}^\textrm{st}$, $\Phi^\textrm{st}=4G x_\textrm{m}^\textrm{st}/\kappa$, is taken into account as well as the reduction of the lattice constant $d$ in presence of atoms for red detuning \cite{Asbo08} 
\begin{equation}
d=\frac{\lambda}{2}\left(1-\frac{\chi^+}{\pi}\right)\,,
\end{equation}
with
\begin{equation}
\chi^+ = \arcsin\left(\frac{\zeta\sqrt{4+\mathcal{A}^2}+\zeta \sqrt{4-\zeta^2\mathcal{A}^2}}{2(1+\zeta^2)}\right)\,.
\end{equation}
The parameter $\mathcal{A}$ quantifies the asymmetry of the lattice
\begin{equation}
\mathcal{A}=\frac{1-R^2}{R} \quad \textrm{with} \quad R=\eta t^2\,.
\end{equation}
Via the equation of motion in Eq.~\ref{eqn:eom} the membrane and atom displacements from the steady state positions $x_\textrm{m}-x_\textrm{m}^\textrm{st}$ and $x_\textrm{i}-x_\textrm{i}^\textrm{st}$ at all later times can be determined. 
Note that for simplicity we will refer to the displacements from the steady state positions as $x_\textrm{m}$ and $x_\textrm{i}$ in the main text of this paper.

\subsection{Measurement setup for back-action measurement}
The measurement setup for the detection of the atomic back-action onto the lattice light is depicted in the inset of Fig.3(a) in the main paper.
 A lock-in amplifier sinusoidally drives the fiber EOM at frequency $\Omega$ imprinting a phase modulation of $\Phi_\textrm{rms}=0.116$ onto the light traveling back to the atoms, which drives the atomic motion. A small fraction of the light is picked up (pick-up reflectivity= 3$\%$) and sent to a photodiode (PD), which records the power modulation due to the atomic back-action. The output voltage of the PD (power-voltage conversion factor of PD=350\,V/W) is measured over the $50\,\Omega$ input resistance of the lock-in with a bandwidth of 18\,Hz. As the fiber EOM reduces the amplitude transmission to $t=0.5$ we operate at a higher power of $P_0=9.12\,$mW. The atom-light detuning is set to $\Delta_\textrm{LA}=-2\pi \times 1\,$GHz for this experiment and the measurement is performed directly after the ramp-up of the lattice.

 The phase shift caused by delays in the measurement setup has been subtracted from the data in Fig.3(b) of the main paper using a reference measurement without atoms, so that only the bare back-action of the atoms onto the lattice light is shown in the figure.

\subsection{Back-action of the atomic ensemble onto the light field}
\subsubsection{One-BS-model}
\begin{figure}%
\includegraphics[width=\columnwidth]{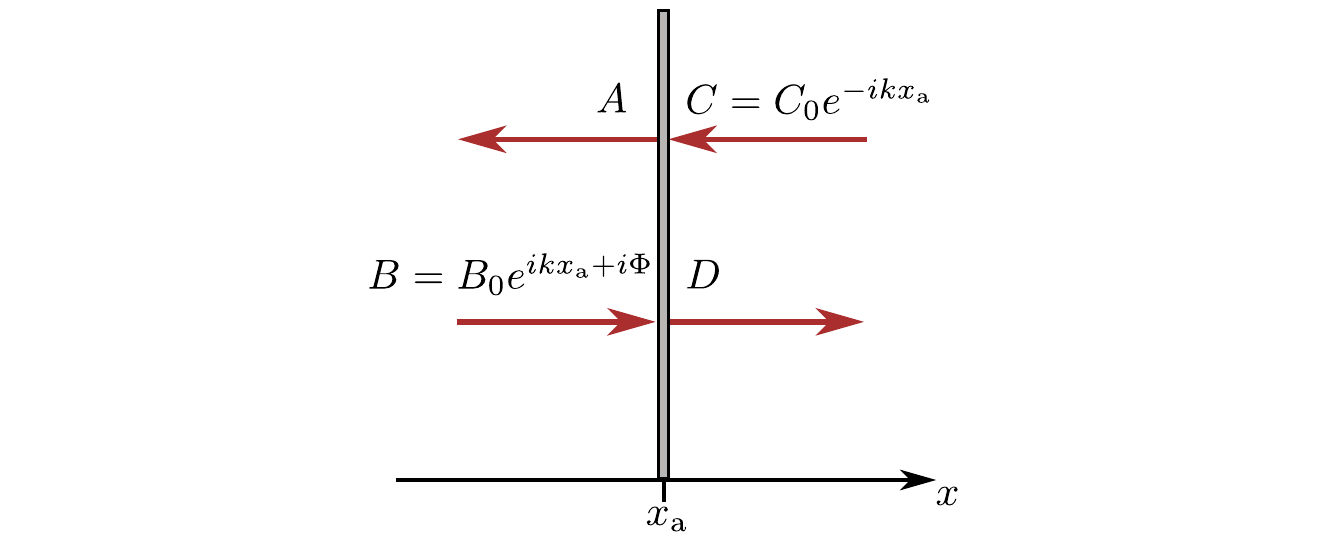}
\caption{{One-BS model with ingoing and outgoing plane waves.}}
\label{fig:Supple2_1BSBackaction}
\end{figure}
To investigate the back-action of a single atomic BS ($N_\textrm{BS}=N$) onto the lattice light we consider a system with one atomic BS and two incident plane waves, one from the right $C(x)=C_0 \exp(-ikx)$ and one from the left $B(x)=B_0 \exp(ikx+i\Phi)$ as depicted in Fig.~\ref{fig:Supple2_1BSBackaction} 
with $B_0=R C_0$. In our back-action experiment the phase $\Phi$ is imprinted onto the light by the fiber EOM.
Using Eqs.~\ref{eqn:MatrixFormulation} and~\ref{eqn:ForceBS} the force on the BS can be calculated.
Absorbing constant terms in a re-definition of the atomic steady state position and in the regime of $k x_\textrm{a},\Phi\ll 1$, we find
\begin{eqnarray}
F_a(x_\textrm{a}) &=&-8k\sigma_\textrm{L}\zeta\frac{\sqrt{I_0I_1}}{c} \left(x_\textrm{a}+\frac{\Phi}{2k}\right)\nonumber\\
&=&-N m\Omega_\textrm{a}^2\left(x_\textrm{a}+\frac{\Phi}{2k}\right)\,,
\end{eqnarray}
where $I_i=\epsilon_0c|E_i|^2/2$, $i \in (0,1)$ are the intensities of the ingoing beams and $k$ is the wavevector.
In presence of an additional damping term, for instance originating in laser cooling, the equation of motion for the position of the BS reads
\begin{equation}
N m \ddot{x}_\textrm{a}=-\Gamma_\textrm{a}N m\dot{x}_\textrm{a}-N m\Omega_\textrm{a}^2\left(x_\textrm{a}+\frac{\Phi}{2k}\right) \,.
\end{equation}
Fourier transforming and solving for $\tilde{x}_\textrm{a}(\Omega)$ gives
\begin{equation}
\label{eqn:Fourierxa}
\tilde{x}_\textrm{a}(\Omega)= -\frac{\Omega_\textrm{a}^2}{2k}\left(\Omega_\textrm{a}^2-\Omega^2+i\Gamma_\textrm{a} \Omega\right)^{-1}\tilde{\Phi}(\Omega)\,.
\end{equation}

From the amplitude $A(x_\textrm{a})$ and $P=\sigma_\textrm{L}\epsilon_0 c|A(x_\textrm{a})|^2/2$ the power modulations $\delta P= P-\langle P \rangle$ of the beam that leaves the atoms towards the left can be calculated
\begin{eqnarray}
\label{eqn:PMod}
\delta P &=& -  4k\sigma_\textrm{L}\zeta \sqrt{I_0I_1}  \left(x_\textrm{a}+\frac{\Phi}{2k}\right)\nonumber\\
&=&-\frac{c}{2}N m\Omega_\textrm{a}^2\left( x_\textrm{a}+\frac{\Phi}{2k}\right)\,.
\end{eqnarray}
Inserting Eq.~(\ref{eqn:Fourierxa}) into the Fourier transform of this expression results in
\begin{equation}
\delta\tilde{P}(\Omega) =-\frac{c}{2}\frac{N m \Omega_\textrm{a}^2}{2 k}\left(1-\frac{\Omega_\textrm{a}^2}{\Omega_\textrm{a}^2-\Omega^2+i\Gamma_\textrm{a} \Omega}\right)\tilde{\Phi}(\Omega)\,.
 \label{eqn:PowerModTheory}
\end{equation}
Eq.~(\ref{eqn:PowerModTheory}) gives the power modulation of beam $A$ caused by a certain modulation of the phase of beam $B$, which is the quantity we measure in the atomic back-action measurement.

\subsubsection{Two-BS-model}
Also for a system of two atomic BSs ($N_\textrm{BS}=N/2$) the back-action onto the lattice light can be calculated.
Referring once more to the multi-BS-system sketched in Fig.~\ref{fig:Supple1_ExactModel}, the power leaving the atomic system towards the left $P=\sigma_\textrm{L}\epsilon_0 c |A_1|^2/2$ can be calculated via the transfer matrix method. Here we treat the case without cavity ($f=1$) and replace the membrane displacement by the corresponding phase shift $x_\textrm{m}=-\Phi/2k$ as we are interested in the power modulation per phase shift. For small atomic displacements from the steady state positions $kx_1,kx_2 \ll 1$ and small phase shifts $\Phi \ll 1$ one finds for the outgoing power modulations $\delta P= P-\langle P \rangle$
\begin{eqnarray}
\delta P&=&4 k \sigma_\textrm{L} \zeta \epsilon_0\frac{c}{2} R  |C_2|^2\left(a_1 x_1+a_2 x_2-a_\Phi \frac{\Phi}{2k}\right)\nonumber \\
&=& \frac{c}{2}N_\textrm{BS}m\Omega_\textrm{a}^2 \left(a_1 x_1+a_2 x_2-a_\Phi \frac{\Phi}{2k}\right)\,,
\end{eqnarray}
with $a_1=-1+9\nu$, $a_2=-1+\nu$ and $a_\Phi=2-10\nu$.
Fourier transforming this expression and combining it with the Fourier transforms of the equations of motion in Eq. (1) of the main paper gives the power modulation of the outgoing beam towards the left ($A_1$) caused by a certain phase modulation of the beam coming from the left ($B_1$)

\begin{eqnarray}
\frac{\delta\tilde{P}(\Omega)}{\tilde{\Phi}(\Omega)} &=&\frac{c}{2}\frac{N m \Omega_\textrm{a}^2}{2 k}\Omega(-i\Gamma_\textrm{a}+\Omega)\nonumber \\
&&\times\frac{(1-5\nu)\Omega(-i\Gamma_\textrm{a}+\Omega)-(-1+\nu)^2 \Omega_\textrm{a}^2}{(\Gamma_\textrm{a}\Omega+i(\Omega^2+(-1+\nu)\Omega_\textrm{a}^2))^2}
\,.
\end{eqnarray}

As we scan the modulation frequency $\Omega$ in the back-action experiment we will excite atoms at different transverse positions in the Gaussian lattice profile with different $\Omega_\textrm{a}$ in contrast to the experiment of Fig.~2 in the main paper, where only atoms resonant with the membrane ($\Omega_\textrm{a}\approx \Omega_\textrm{m}$) are driven.
Therefore, for modeling the back-action we take into account all atoms ($N=N_\textrm{lat}$) and use an increased, inhomogenously broadened linewidth $\Gamma_\textrm{a}$ much larger than the laser-cooling rate. A more precise modeling of the atomic ensemble is not possible as we lack information on the exact properties of the atoms in the lattice. The simplified model does not allow us to exactly reproduce the measured back-action, but it clearly shows a significant difference between one- and two-BS model in the phase behavior of the back-action for realistic parameters as depicted in Fig.~3 of the main paper.  
  
\begin{figure}%
\includegraphics[width=\columnwidth]{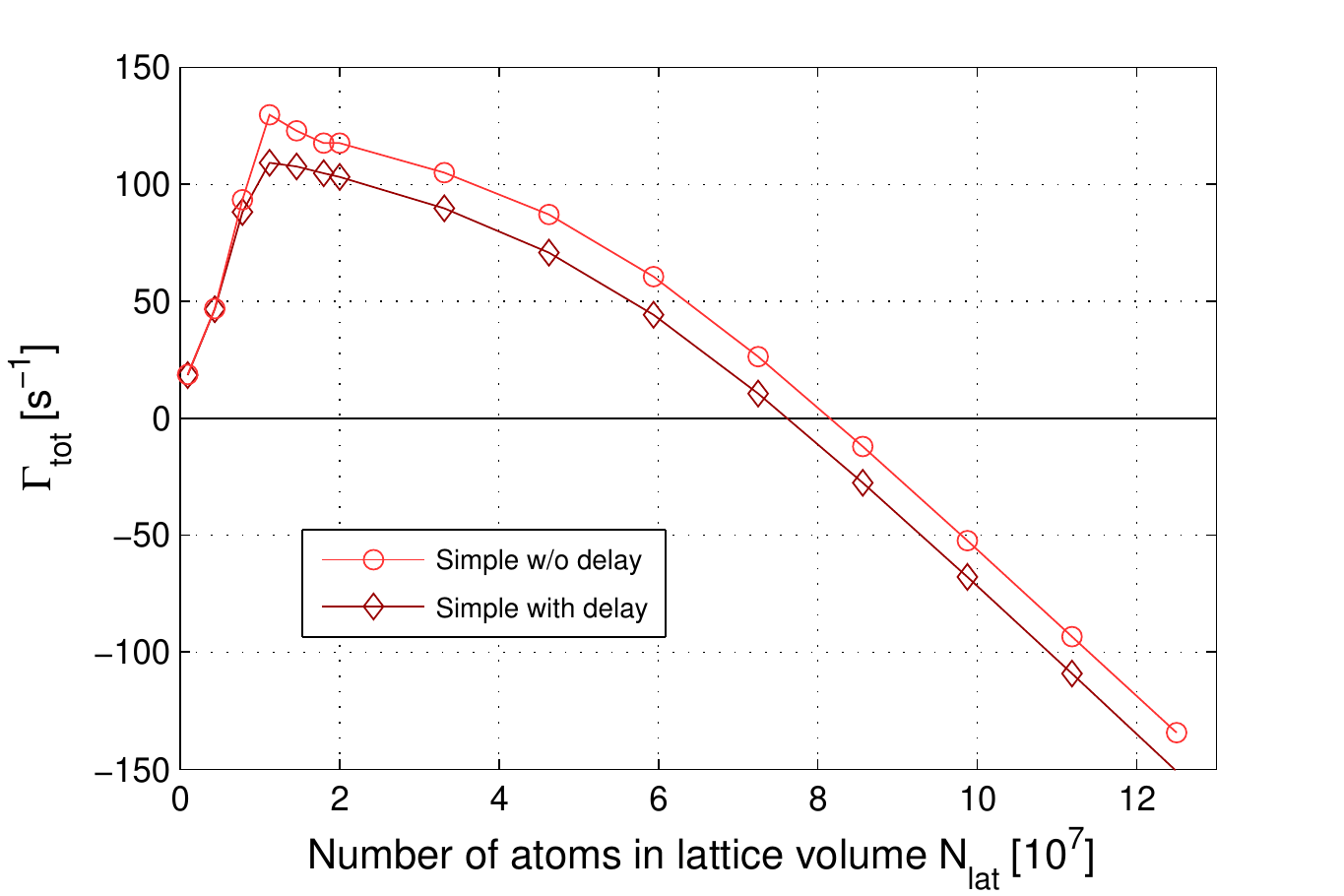}
\caption{{Effect of the delay. $\Gamma_\textrm{tot}$ extracted from numerical simulation of Eq.~\ref{eqn:EOMwithDelay}} with $\tau=36\,$ns (dark red diamonds) and $\tau=0\,$ns (light red circles). All other parameters are listed in the main text of the paper.}
\label{fig:Supple3_WithDelay}
\end{figure}

\subsection{Linearized two-BS model with delay}

To investigate the effect of the propagation time delay between atoms and membrane $\tau_\textrm{prop}=30\,$ns and the finite cavity response time $\tau_\textrm{cav}=0.6\,$ns on the coupled dynamics we insert the total retardation of $\tau=\tau_\textrm{prop}+\tau_\textrm{cav}$ into the forces from the atoms onto the membrane and vice versa. The equations of motion presented in Eq. (1) in the main paper but with the retardation included are
\begin{widetext}
\begin{eqnarray}
\label{eqn:EOMwithDelay}
\ddot{x}_\textrm{m}(t)&=&-(\Gamma_\textrm{m}+\Gamma_\textrm{opt})\dot{x}_\textrm{m}(t)-\Omega_\textrm{m}^2 x_\textrm{m}(t)+k_\textrm{mm}x_\textrm{m}(t)+k_\textrm{m1}x_\textrm{1}(t-\tau)+k_\textrm{m2}x_\textrm{2}(t-\tau)\,, \nonumber \\
\ddot{x}_\textrm{1}(t)&=&-\Gamma_\textrm{a}\dot{x}_\textrm{1}(t)+k_\textrm{1m}x_\textrm{m}(t-\tau)+k_\textrm{11}x_\textrm{1}(t)+k_\textrm{12}x_\textrm{2}(t)\,, \nonumber \\
\ddot{x}_\textrm{2}(t)&=&-\Gamma_\textrm{a}\dot{x}_\textrm{2}(t)+k_\textrm{2m}x_\textrm{m}(t-\tau)+k_\textrm{21}x_\textrm{1}(t)+k_\textrm{22}x_\textrm{2}(t)\,,
\end{eqnarray}
\end{widetext}
with the coefficients $k_{ij}$ $(i,j\in[m,1,2])$ as in Eq. (2) of the main paper.

Fig.~\ref{fig:Supple3_WithDelay} shows the total membrane damping rate $\Gamma_\textrm{tot}$ extracted from numerical simulation of Eq.~\ref{eqn:EOMwithDelay} in presence and absence of the delay for different atom numbers. The curve without delay is a copy of the curve from red empty triangles of Fig. 2(b) in the main paper. The curve with delay has been evaluated with for the same parameters but $\tau=36\,$ns. The presence of the delay slightly decreases $\Gamma_\textrm{tot}$ and the threshold atom number at which $\Gamma_\textrm{tot}$ becomes negative. However, even if the delay in the simulation is slightly larger than the measured delay, the effect is small. This is in agreement with our measurements, which did not show a significant modification of the behavior when we changed the path length between atoms and membrane.

\end{document}